\documentclass[aps,pra,groupedaddress,showpacs, preprint]{revtex4}

\usepackage{epsfig}
\usepackage{color}
\begin{document}

\title{Two-step Doppler cooling of a three-level ladder system with an intermediate metastable level}

\author{C. Champenois}
\email{caroline.champenois@univ-provence.fr}
\author{G. Hagel}
\author{M. Knoop}
\author{M. Houssin}
\author{C. Zumsteg}
\author{F. Vedel} \affiliation{Physique
des Interactions Ioniques et Mol\'eculaires (CNRS UMR 6633),
Universit\'e de Provence, Centre de Saint J\'er\^ome, Case C21,
13397 Marseille Cedex 20, France}
 \author{M. Drewsen}
\affiliation{QUANTOP, Danish National Research foundation Center for Quantum Optics,
Department of Physics and Astronomy, University of Aarhus, DK-8000 {\AA}rhus C., Denmark} 
\date{\today}

\begin{abstract} 
Doppler laser cooling of a three-level ladder system using two near-resonant laser fields is analyzed in the case of the intermediate level being metastable while the upper level is short-lived. Analytical as well as numerical results for e.g. obtainable scattering rates and achievable temperatures are presented. When appropriate, comparisons with  two-level single photon Doppler laser cooling is made. These results are relevant to recent experimental Doppler laser cooling investigations addressing intercombination lines  in alkali-earth metal atoms and quadrupole transitions in alkali-earth metal ions.    
\end{abstract}


\maketitle

\section{introduction}
Today, laser cooling of atoms is widely used in experiments where high precision spectroscopy or precise control of the atomic motion is required. A large variety of schemes has been developed and applied to suit atoms with specific level structures and for obtaining particular temperature ranges \cite{metcalfbook}. In general, the lower the wanted temperature, the more sensitive does the light scattering process has to be on the velocity of the atom. For the most simple laser cooling scheme relying on the Doppler shift of an optical transition, Doppler laser cooling, this means the narrower the linewidth of the optical transition, the lower the obtainable temperature. However, since the maximum cooling force in the Doppler cooling scheme is dependent on the photon scattering rate, narrow linewidth transitions will lead to longer cooling times than wider transitions. 

The aim of the present study is to show that by driving, with two step-wise near-resonant laser fields, a three-level ladder system having  an intermediate metastable level and a dipole emitting upper level, it is indeed possible to obtain both short cooling times and at the same time low temperatures through changes of the laser parameters. The present results  concern in particular  
alkali-like ions  Ba$^+$, Sr$^+$, or Ca$^+$ \cite{hendricks07}, but also contain information relevant for Doppler cooling of alkali-earth metal atoms using intercombination lines~\cite{curtis01, binnewies01}. In addition to the excellent laser cooling properties, the presented scheme has the major advantage that individual ions can be detected by dipole emitted light from the upper level to the ground level at a wavelength being the sum of those of the two excitation fields. By the appropriate choice of laser parameters, a large scattering rate can be obtained, which can lead to an essentially background-free detection as it has been  demonstrated experimentally recently  \cite{hendricks07}.

Previously, ladder- (or cascade-) schemes have been  studied
in~\cite{magno03, malossi05,morigi07, dunn07} in the prospect of two-photon two-color Doppler cooling on narrow lines like demonstrated experimentally in~\cite{malossi05} on alkaline-earth metals using an intercombination line. In these studies, the excited state of a laser-driven transition is coupled by a second laser to an upper more stable level. In the present study, the  excited metastable level is coupled to a higher short-lived level, to enhance its coupling with the ground state. Such an excitation scheme is used for example for sideband cooling, but while sideband cooling processes have already been studied theoretically~\cite{marzoli94} and experimentally~\cite{diedrich89, roos99} in the low saturation regime, the purpose of the present study is to go beyond the low saturation regime and to consider Doppler cooling in the regime where the sidebands are not resolved, as considered experimentally previously~\cite{hendricks07}. Our analysis applies to any atomic system that can be described by the ladder scheme depicted on Fig.~\ref{fig_ladder}. Nevertheless, when numerical applications are needed, we refer to parameters describing the alkali-like ions  Ba$^+$, Sr$^+$, or Ca$^+$.

This article is organized as follows.  Sec.~\ref{model} introduces the
theoretical model used to predict the population of the internal states. In Sec.~\ref{immobile} the
response of an atom at rest to the lasers excitation is studied, focussing on the expected signal.  
 The Doppler cooling efficiency of the proposed excitation scheme is studied in Sec.~\ref{cooling}, and in Sec.~\ref{Sec:Conclusions}
conclusions are reported.

\section{Theoretical model} \label{model}

\begin{figure}[htb]
\begin{center}
\epsfig{file=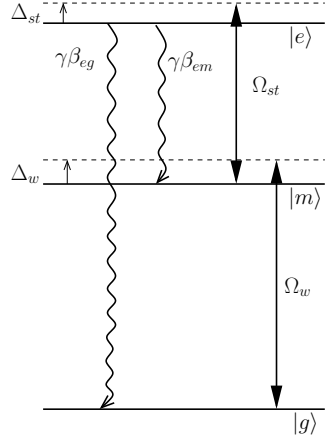, height=0.35\hsize} \caption{Ladder level scheme. The straight arrows represent the laser coupling between states $|g\rangle$, $|m\rangle$ and $|e\rangle$.  The wavy lines indicate radiative decay taken into account in the model.} \label{fig_ladder}
\end{center}
\end{figure}

The considered atomic system is composed by three electronic levels
which are coupled by laser fields, according to the ladder-scheme
depicted in Fig.~\ref{fig_ladder}. State $|g\rangle$ is the ground state, $|m\rangle$ is metastable, and the excited  state $|e\rangle$ decays radiatively into $|g\rangle$ and $|m\rangle$.
This configuration of levels occurs, for instance, in
alkaline-earth atoms such as Sr and Ca, rare-earth atoms like Yb and in alkali-like
ions with a metastable $d$-orbital, such as Hg$^+$, Yb$^+$, Ba$^+$,
Sr$^+$, or Ca$^+$.  In these last cases, the
relevant levels can be identified with the states
$|g\rangle=|S_{1/2}\rangle$, $|m\rangle=|D_{3/2}\rangle$ and $|e\rangle=|P_{1/2}\rangle$.
Then, the transition $|g\rangle\to|m\rangle$ is an electric quadrupole transition with a
linewidth in the range from 0.01 to 0.1~Hz, while $|g\rangle$ and
$|m\rangle$ couple to the excited state $|e\rangle$ with electric
dipole transitions. We denote by
$\omega_{ij}$ ($i,j=g,m,e$) the resonance frequencies of the
transitions. Transitions $|g\rangle\to |m\rangle$, marked by $w$ (weak), and
$|m\rangle\to |e\rangle$, marked by $st$ (strong) are coupled by lasers at frequency
$\omega_w$ and $\omega_{st}$, the strengths of the couplings are characterized by the Rabi frequencies $\Omega_w$ and $\Omega_{st}$ and the detunings defined as $\Delta_{w}=\omega_w-\omega_{mg}$, and
$\Delta_{st}=\omega_{st}-\omega_{em}$.  The radiative process
coupling $|e\rangle$ to states $|g\rangle$ and $|m\rangle$ is characterized by the decay rate $\gamma$ and the
branching ratio $\beta_{eg}/\beta_{em}$ ($\beta_{em}+\beta_{eg}=1$). The radiative decay of state  $|m\rangle$, whose natural lifetime is of the order
of 1 s for the mentioned ions, is neglected in the analytical model we
present below. 

We denote by $\rho$ the density matrix for the atom's internal
degrees of freedom, while we treat the center of mass variables
classically. $x(t)$ denotes the time-dependent position of
the atom. The master equation for the density matrix $\rho$ is
\begin{equation} \label{eq_master}
 \frac{\partial}{\partial
t}\rho=-\frac{\rm i}{\hbar}[H,\rho]+{\cal L}\rho \end{equation}
where Hamiltonian $H$ gives the coherent dynamics and is
decomposed into the terms 
\begin{eqnarray}
H=H_0+H_I
 \end{eqnarray}
 where \begin{eqnarray} H_0
&=&-\hbar\Delta_w|m\rangle\langle m|-\hbar(\Delta_w+\Delta_{st})|e\rangle\langle
e| 
 \end{eqnarray}
gives the internal energies in the reference frames of the lasers
 while
 \begin{eqnarray} \label{Ham}H_I
&=&\frac{\hbar\Omega_w}{2}{\rm e}^{{\rm i}k_wx(t)}|m\rangle\langle
g|+\frac{\hbar\Omega_{st}}{2}{\rm e}^{{\rm i}k_{st}x(t)}|e\rangle\langle m|+{\rm H.c.}  \end{eqnarray}
gives the laser coupling, where $k_j$ denotes the  laser wave vectors
($j=w,st$). The relaxation operator
 \begin{eqnarray}
 {\cal L}\rho
&=&-\frac{1}{2}\gamma \left(\rho|e\rangle\langle
e|+|e\rangle\langle e|\rho\right) \nonumber \\
& &+\beta_{eg}\gamma|g\rangle\langle e|\rho|e\rangle\langle g|
+\beta_{em}\gamma|m\rangle\langle e|\rho|e\rangle\langle m|
 \end{eqnarray} describes the radiative processes.

The response of the atomic system to laser excitation is revealed by
  the occupation probabilities of the atomic levels, which
can be monitored by the number of scattered photons.  Detection of the atoms is made on the transition with the highest fluorescence rate  $|e\rangle\to|g\rangle$ (branching ratio $\beta_{eg}/\beta_{em}>1$ for the ions mentioned above) which lies in the blue domain ($< 500$ nm). For the excitation scheme discussed in the following, this allows for photon detection without scattered light background since the excitation wavelengths lie in the red and near infra-red domain ($>680$ nm). This major advantage does not hold for Hg$^{+}$ nor Yb$^+$ for which the excitation and emission wavelengths are very close. In the following, we focus on the case of Ba$^+$, Sr$^+$ and Ca$^+$ for the choice of the relevant parameters needed to describe the atomic system.

It should  be noted that the dipole (strong) and
the quadrupole (weak) couplings differ  by some orders of magnitude. Nevertheless, powerful lasers in the red and near infra-red domain are now widely available and, assuming proper focussing of the laser beam,  Rabi frequencies of the order of 1 MHz can be reached on the quadrupole transition.  To give an idea of the feasible Rabi frequencies, let us mention that, for Ca$^+$, 200 mW of laser power at 729~nm focussed on a waist radius of 50 $\mu$m results in $\Omega_w \simeq 1$ MHz ~\cite{hendricks07}.

\section{Internal dynamics of a localized
particle\label{immobile}}

In this section we look for the solutions of Eq.~(\ref{eq_master})
when the motion of the particle can be neglected.  For numerical applications and when approximations are needed in analytical expressions,  parameters leading to a strong driving of the  $|m\rangle \to |e\rangle$ transition with  reduced sensitivity to Doppler shift are chosen ($\gamma^2 \ll \Omega_{st}^2, \Delta_{st}^2$). Furthermore, to remain close to experimental conditions, we do not require excessively large Rabi frequency on the   $|g\rangle \to |m\rangle$ transition and assume $\Omega_w^2 \ll \gamma^2, \Omega_{st}^2$.

\begin{figure}[htb]
\begin{center}
\epsfig{file=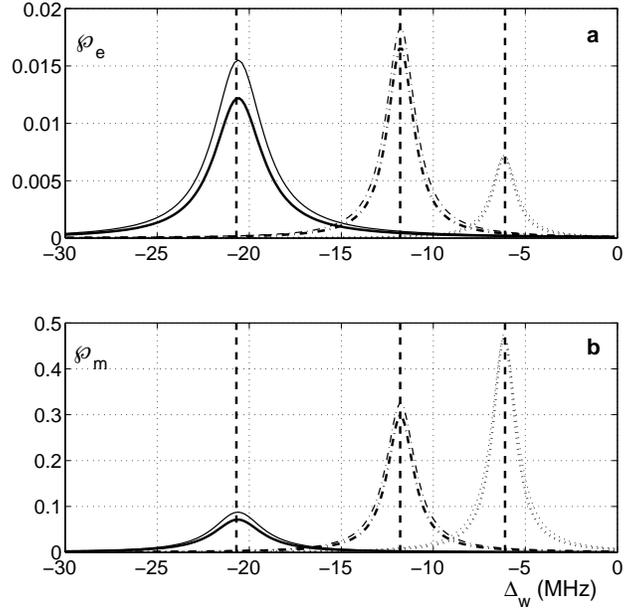, height=0.5\hsize} \caption{Steady state population of level  $|e\rangle$ ({\bf a}) and  $|m\rangle$ ({\bf b}) versus $\Delta_{w}$ for $\gamma/2\pi=20$ MHz, $\Omega_{st}/2\pi=$ 100 MHz and $\Omega_w/2\pi=$ 1 MHz. $\Delta_{st}/2\pi=-100$ MHz (solid line), $\Delta_{st}/2\pi=-200$ MHz (dot-dashed line), $\Delta_{st}/2\pi=-400$ MHz (dashed line). In each case, the thick lines give result for $\beta_{eg}=1$ and the narrow line for $\beta_{eg}=0.75$. The vertical dashed lines show the light-shifts $\delta_{LS}$ calculated for the three sets of curves.  } \label{fig_popLS}
\end{center}
\end{figure}

In order to get some insight in the problem,  the dressed state picture of
the system  can be of help in the limiting case where the weak coupling beam can be considered as a probe of the atom dressed by the photons of the strong transition. In this picture, the $|m\rangle$ state is broadened and shifted by the strong coupling $\Omega_{st}$. For negative detuning $\Delta_{st}$  we expect the $|g\rangle \to |m\rangle$ transition resonance to be shifted to $\delta_{LS}=-(\sqrt{\Omega_{st}^2+\Delta_{st}^2}+\Delta_{st})/2$. This is confirmed on Fig.~\ref{fig_popLS} where the lineshape of the excitation probability in the $|m\rangle$ state  $\wp_m=\langle m|\rho|m \rangle$ shows a maximum  very close to this value. Increasing  the detuning $\Delta_{st}$ decreases the light-shift as well as  the  power-broadened linewidth of the  $|g\rangle \to |m \rangle$ transition, resulting in an increase of the maximum  of $\wp_m$ for constant coupling strength $\Omega_{st}$. 

In an experiment, the measured signal is proportional to the  probability $\wp_e=\langle e|\rho|e \rangle$
 to be in the $|e\rangle$ state. Its dependance with the detuning $\Delta_{st}$ does not follow the same rule as $\wp_m$ and Fig.~\ref{fig_popLS} clearly shows that a maximum excitation probability can be reached for a good choice of this detuning. This behaviour is studied in the following through resolution of the optical Bloch equations resulting from the master equation   Eq.~(\ref{eq_master})

\subsection{Maximum fluorescence rate\label{sec_popP}}

The steady state solution of Eq.~(\ref{eq_master}) can be written analytically, assuming conservation of the total population ($\wp_g+\wp_m+\wp_e=1$) and without any approximation made.  For the population $\wp_e$, this solution can be written like 
\begin{equation}\label{eq_popP}
\wp_e= \Omega_{st}^2 \Omega_w^2/ {\mathcal D}
\end{equation}
with
\begin{eqnarray}
{\mathcal D} & = & \beta_{eg}\left(4\Delta_w(\Delta_w+\Delta_{st})\left(4\Delta_w(\Delta_w+\Delta_{st})+2\Omega_w^2-2\Omega_{st}^2\right)+4\Delta_w^2 \gamma^2 +\Omega_{st}^4 \right) \nonumber \\
          &  & +(\beta_{eg}-1)2\Omega_w^2(2\Delta_w^2+4\Delta_w\Delta_{st}-\Omega_{st}^2)+2\Omega_w^2\left(4(\Delta_w+\Delta_{st})^2+\gamma^2+\Omega_w^2\right).   
\end{eqnarray}

This equation is used to estimate the maximum scattering rate that can be expected from this excitation scheme, the effective linewidth of the forbidden transition once coupled by the strong laser excitation, and the dependence of these two entities on the laser coupling parameters. Despite its complexity, Eq.~\ref{eq_popP} can be connected to the dressed state picture if one keeps in mind that for $\Delta_w=\delta_{LS}$, $4\Delta_w(\Delta_w+\Delta_{st})=\Omega_{st}^2$.

 In the limit where $\Omega_w^2 \ll \gamma^2 |\Delta_w/\Delta_{st}|$, the condition for maximum occupation probability of the $|e\rangle$ state, and  thus for maximum fluorescence signal, can also be derived analytically from  Eq.~\ref{eq_popP}. In  first order in  $\gamma^2/(\Omega_{st}^2+\Delta_{st}^2)$, this maximum  is found for 
\begin{equation}\label{eq_corrLS}
\Delta_w=\delta_{LS}\left(1-\frac{\gamma^2}{4(\Omega_{st}^2+\Delta_{st}^2)}\right).
\end{equation} 
This correction to the light-shift can be omitted in the evaluation of the maximum population $\wp_e^M$  as it adds a correction  in second order in  $\gamma^2/(\Omega_{st}^2+\Delta_{st}^2)$, which is negligible  for the set of parameters of interest. For $\Omega_w^2 \ll \Omega_{st}^2$, this maximum is
\begin{equation}\label{eq_maxP}
\wp_e^M=1\Big/ \left( 4+\frac{2\Delta_{st}}{\delta_{LS}}+\beta_{eg}\frac{4 \delta_{LS}^2 \gamma^2}{\Omega_{st}^2 \Omega_w^2}+(\beta_{eg}-1)\frac{4\delta_{LS}(\delta_{LS}+2\Delta_{st})}{\Omega_{st}^2}+\frac{2\gamma^2}{\Omega_{st}^2}\right).
\end{equation}
This last equation rules the maximum scattering rate, observed when the detuning of the weak transition is adjusted to the light-shift induced by the strong coupling. Assuming this detuning $\Delta_w$ is adjusted, for each Rabi frequency $\Omega_w$ (which is experimentally the limiting factor) there is an optimal   set  ($\Omega_{st}, \Delta_{st}$) which maximises $\wp_e^M$. This optimal set can be found numerically but better understanding  can be gained in the particular case of large detunings  $\Delta_{st}^2 \gg \Omega_{st}^2  $ for which the light-shift can be approximated by the simple form  $\delta_{LS}=\Omega_{st}^2/4\Delta_{st}$. With this simplification, the optimal set is described by the analytic expression:
\begin{equation}\label{eq_opt}
\left(\frac{\Omega_{st}}{\Delta_{st}}\right)^2_{opt}=4\frac{\Omega_w}{\gamma} \sqrt{\frac{2}{\beta_{eg}}},
\end{equation}
which corresponds to  the maximum population 
\begin{equation}\label{eq_Popt1}
\wp_e^M|_{opt} = \frac{1}{4+2(\beta_{eg}-1)+2\sqrt{2\beta_{eg}}\gamma/\Omega_w+2\gamma^2/\Omega_{st}^2}.
\end{equation}
This expression is simplified in 
\begin{equation}\label{eq_Popt}
\wp_e^M|_{opt}\simeq\frac{\Omega_w}{\gamma }\frac{1}{2 \sqrt{2\beta_{eg}}},
\end{equation}
if $\gamma^2/\Omega_{st}^2$ is negligible compared to $\gamma/\Omega_{w}$.

Equation (\ref{eq_Popt}) confirms the dependance of the maximum population $\wp_e^M$ with $\beta_{eg}$, observed on Fig.~\ref{fig_popLS}. Indeed, this population increases when the branching factor $\beta_{eg}$ decreases. For the ions considered in this work, this factor is of the order of 0.75 for Ba$^+$ and 0.95 for Ca$^+$ and Sr$^+$. Since the previous equations show that different values of $\beta_{eg}$  result only in a variation of the maximum population (Eq.~\ref{eq_Popt})  and optimal set of parameters to reach it (Eq.~\ref{eq_opt}), $\beta_{eg}$ is fixed to 1 in the following. This simplification results mainly in a slight underestimation of the number of scattered photons but allows to handle more general equations.

\begin{figure}[htb]
\begin{center}
\epsfig{file=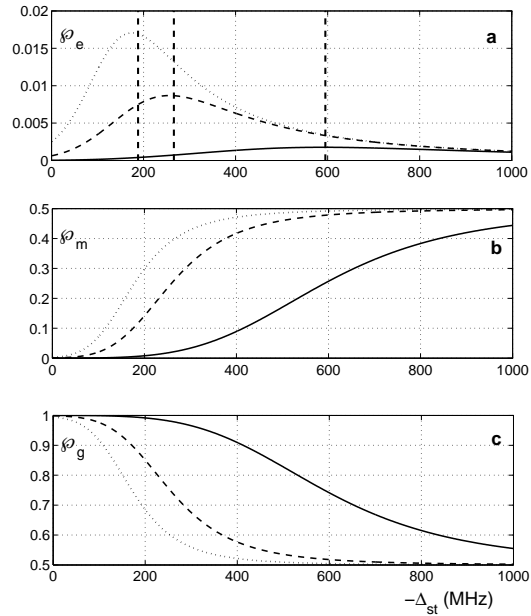, height=0.5\hsize} \caption{Steady state population of level  $|e\rangle$  ({\bf a}),  $|m\rangle$  ({\bf b}), $|g\rangle$  ({\bf c}) versus $-\Delta_{st}$ for an atomic system close to  Ba$^+$, Sr$^+$, or Ca$^+$ : $\gamma/2\pi=20$ MHz, in the simplified case where $\beta_{eg}=1$. The different curves are for different values of $\Omega_w/2\pi$: 0.1 MHz (solid line), 0.5 MHz (dashed line) and 1 MHz (dotted line).  $\Omega_{st}/2\pi= 100$ MHz for all curves and   $\Delta_w$ is adjusted to $\delta_{LS}$ for each value of $\Delta_{st}$.  The vertical dashed lines show the optimum detunings like defined by Eq.~\ref{eq_opt}, for the three sets of curves.    } \label{fig_optP}
\end{center}
\end{figure}

Beside the dependance on the branching factors, it is worth noting that the optimal population given in Eq.~\ref{eq_Popt} behaves like $\Omega_w/\gamma$, and not like $\Omega_w^2/\gamma^2$, as a simple two level picture would have suggested. This linear behaviour is confirmed by numerical calculations, even out of the range of the simplifying approximation  $\Omega_{st}^2 \ll \Delta_{st}^2$. This is illustrated on Fig.~\ref{fig_optP} where the probabilities of occupation of the three levels,   derived from Eq.~\ref{eq_master} with no approximations, are plotted for different values of $\Omega_w$. For each detuning on the strong transition $\Delta_{st}$, the weak transition detuning is adjusted to the light-shift due to the strong coupling: $\Delta_w=\delta_{LS}$. For a given value of $\Omega_{st}$, these curves confirm an optimum value for $\Delta_{st}$ which maximises the resonance value   $\wp_e^M$. The numerical results also confirm the linear dependance of the maximal population with $\Omega_w$ and obey the simplified analytic solution given in Eq.~\ref{eq_Popt} to better than  3\%. Following Eq.~\ref{eq_opt}, the optimum of $\Delta_{st}$ is expected to behave like $1/\sqrt{\Omega_w}$. This is also confirmed by the numerical results which show that the simplified equation Eq.~\ref{eq_opt} reproduces numerical observation for optimal $\Delta_{st}$ to better than  7\% for the value of $\Omega_w$ chosen for this figure.

% Fig.\ref{fig_optP} clearly presents evidence of the existence of a maximum probability $\wp_e$, which corresponds to a maximum scattering rate, for an optimum laser detuning $\Delta_{st}$. This maximum of $\wp_e$ do%es not correspond to a maximum saturation of the $|g\rangle \to | m\rangle$ transition and an equivalent maximum can not be found in the $|m\rangle$ population.

We can compare this result with the case where the strong coupling is maximum ($\Delta_{st}=0$).   Then $\delta_{LS}=\pm \Omega_{st}/2$ and in the regime described by $\Omega_{st}>\gamma/2$, the maximum of $\wp_e^M$  is $(\Omega_w/\gamma)^2/\beta_{eg}$. For the realistic conditions we are concerned with (namely $\Omega_w<\gamma$), the maximum population in $|e\rangle$ and thus the maximum number of scattered photons is smaller by a factor proportional to $\gamma/\Omega_w$ to what can be expected for large detuning $\Delta_{st}$ (see  Eq.~\ref{eq_Popt} for comparison).

To know if the ladder-scheme  can result in enough scattered photons to be efficiently detected, it is useful  to compare this scheme with the one usually used for  detection and Doppler cooling  of the considered trapped ions \cite{neuhauser80}. It consists in the  excitation on the $|g\rangle \to |e\rangle$  and   $|m\rangle \to |e\rangle$ transitions, the last one being required for repumping ions decayed in the metastable state. This system forms a $\Lambda$-scheme and solving the optical Bloch equations for this system,  one can show  that the maximum occupation probability in the $|e\rangle$ state is close to 20 \%. The parameters chosen for Fig.~\ref{fig_optP} show a maximum close to 2 \% for the ladder-scheme. The complete absence of background signal due to scattered light largely compensates a reduction of absolute fluorescence signal by an order of magnitude. Furthermore, an occupation probability of the excited state as high as 10 \% can be found by numerical calculations with  Rabi frequencies on the weak transition  of  7 MHz. These are achieved by tight focussing of laser power superior to 500 mW.

\subsection{Resonance linewidth\label{sec_width}}

\begin{figure}[htb]
\begin{center}
\epsfig{file=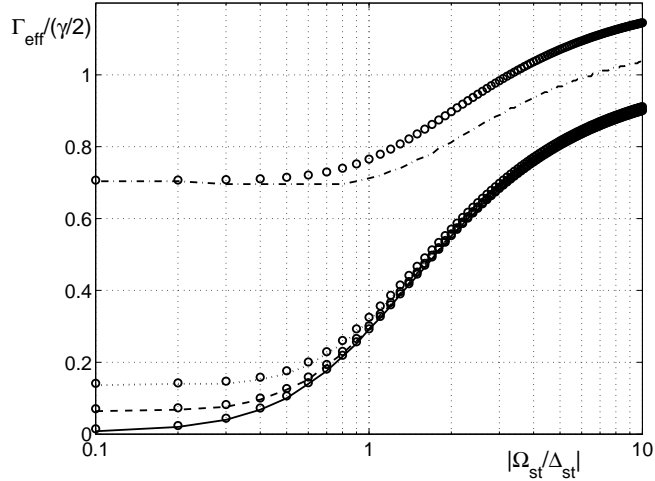, height=0.4\hsize} \caption{FWHM of the resonance curve of $\wp_e$ with $\Delta_w$, in units of $\gamma/2$, as a function of the ratio $|\Omega_{st}/\Delta_{st}|$, for a system identical to Fig.~\ref{fig_optP}. $\Delta_{st}/2\pi=-200$ MHz and $\Omega_{st}$ varies from $0.1|\Delta_{st}|$ to $10|\Delta_{st}|$. The different curves are for different values of $\Omega_w$ which are 0.1 MHz (solid line), 0.5 MHz (dashed line), 1 MHz (dotted line) and 5 MHz (dot-dashed line). The lines give the numerical results from  the master equation, with no approximations. The symbols plots the FWHM as given by Eq.~\ref{eq_Geff}.  } \label{fig_widthP}
\end{center}
\end{figure}

The linewidth $\Gamma_{eff}$ of the fluorescence resonance profile with the laser detuning $\Delta_w$ and its dependance on the two couplings can be also studied numerically.  This study gives evidence for two different types of behaviour for  this linewidth (see Fig.~\ref{fig_widthP}): When $\Omega_{st}$ is small compared to $\Delta_{st}$, the linewidth has a constant value, which increases linearly with $\Omega_w$. Once  $\Omega_{st} > \Delta_{st}$, the linewidth grows with $\Omega_{st}$ and becomes independent on $\Omega_w$. To reflect this behaviour in the analytic expression of the resonance linewidth, we first assume that $\Omega_w$ is the smallest parameter of the problem ($\Omega_w^2 \ll \Omega_{st}^2, \Delta_{st}^2, \gamma^2, \Delta_w^2$), to extract a limit linewidth we call $\Gamma_0$ (FWHM):
\begin{equation}
\Gamma_{0}=\frac{2 \delta_{LS}}{1+4(\Omega_{st}^2+\Delta_{st}^2)/\gamma^2}\sqrt{2+4(\Omega_{st}^2+\Delta_{st}^2)/\gamma^2}.
\end{equation}
This expression is simplified in the regime where $\gamma^2 \ll \Omega_{st}^2+\Delta_{st}^2$ and can be written like: 
\begin{equation}\label{eq_width}
\Gamma_{0}\simeq \frac{\gamma}{2}\left(1-\frac{1}{\sqrt{1+\Omega_{st}^2/\Delta_{st}^2}}\right).
\end{equation}
Then, we assume that  the increasing  weak transition coupling  power broadens this limit linewidth, resulting in  the effective linewidth $\Gamma_{eff}$ 
\begin{equation}\label{eq_Geff}
\Gamma_{eff}=\sqrt{\Gamma_0^2+2\Omega_w^2}.
\end{equation}
For strong saturation ($\Omega_{st}^2/\Delta_{st}^2 \gg 1$), $\Gamma_0 \simeq \gamma/2(1-\Delta_{st}/\Omega_{st})$ tends to the maximum value of $ \gamma/2$, and $\Gamma_{eff} \simeq \Gamma_0$. For low saturation, the limit linewidth behaves like $\Gamma_0 \simeq \gamma/2(\Omega_{st}^2/2\Delta_{st}^2)$ and tends to 0, the main contribution to the linewidth is then due to the weak coupling: $\Gamma_{eff} \simeq \sqrt{2}\Omega_w$.

This dependance of $\Gamma_{eff}$ with $|\Omega_{st}/\Delta_{st}|$ is shown on Fig.~\ref{fig_widthP} for different values of $\Omega_w$ and a far detuned strong coupling laser $\Delta_{st}=-10 \gamma$. The comparison between Eq.~\ref{eq_Geff} and the exact FWHM of the resonance profile of $\wp_e$ with $\Delta_w$, calculated numerically, shows a very good agreement as long as $\Omega_w$ remains weak compared to $\gamma/2$. When this approximation is not valid, the methods used above result in an overestimation of the  linewidth for $\Omega_{st} \geq \Delta_{st}$ (dash-dotted line of Fig.~\ref{fig_widthP}). These curves show that the effective linewidth of the excitation scheme can be adjusted continuously by the ratio  $|\Omega_{st}/\Delta_{st}|$. When this ratio is chosen to optimize the maximum of $\wp_e$ (see Eq.~\ref{eq_opt}), the effective linewidth can be evaluated through Eq.~\ref{eq_Geff}. For $\Omega_w/\gamma \ll 1$, $\Gamma_0 \simeq \sqrt{2} \Omega_w$ and $\Gamma_{eff}\simeq 2 \Omega_w$. This value gives the lower boundary for an estimation of the effective linewidth when parameters are set to optimise the excited population.

\section{Two-step Doppler cooling\label{cooling}}

So far we have neglected the effect of the center-of-mass motion on the atomic
dynamics. In the following,  the considered atoms are trapped in a harmonic potential, like in \cite{hendricks07}, their motion is periodic  and can give rise to sidebands \cite{wineland79}. To quantify the efficiency of the proposed two-step Doppler cooling, we  assume that the effective linewidth $\Gamma_{eff}$ is larger than the frequencies of oscillation $\omega_i$ of the ions in the trap. Practically, it means that the sidebands are not resolved in the absorption spectra and from the point of view of laser-atom interactions, the atoms  can be considered as free \cite{wineland79}. Furthermore, we also assume that the recoil frequency $\hbar k_x^2/2m$ ($x$ being $w$ or $st$) is smaller than the effective linewidth \cite{dalibard84} and that the frequencies of oscillation are high enough so that the root mean square of the velocity $v_{rms}$ is not significantly changed during one oscillation period. Within these boundary conditions, the effect of the atom motion can be taken into account in the master equation (Eq.~\ref{eq_master}) and its stationary state solution (Eq.~\ref{eq_popP}) by the Doppler effect which shifts the detuning from $\Delta_{st}$ to $\Delta_{st}-\mathbf{k}_{st}.\mathbf{v}$ and $\Delta_w$ to $ \Delta_w-\mathbf{k}_w.\mathbf{v}$.

To simplify the notations, we consider a one dimensional problem, extrapolation to three dimensions being obvious.  We assume that the two lasers propagate along the same axis, so $v$ is the velocity along that axis and the positive sign is defined with respect to $\mathbf{k}_w$. Consequently, the projection $k_{st}$ is positive (negative) if the laser on the strong transition copropagates (counterpropagates) with the laser on the weak transition. By definition, $k_w$ is always positive.

For  travelling waves,  the radiative force $F$ leading to Doppler cooling results from  momentum transferred to the atom by photons on  absorption/emission cycles. It is exactly the gradient of the atom-light interaction potential  \cite{dalibard84} given in Eq.~\ref{Ham}: $F=\hbar k_w \Omega_w {\rm Im}(\rho_{gm})+\hbar k_{st} \Omega_{st} {\rm Im}(\rho_{me})$. Using the relations induced for the stationary state density matrix by the master equation, one can show that this definition is equivalent to $F(v)=\hbar k_w \beta_{eg}\gamma \wp_e(v)+\hbar k_{st}\gamma \wp_e(v)$. In the simplifying case where the decay  from $|e\rangle$ to $ |m\rangle$ is neglected ($\beta_{eg}=1$), the radiative force is 
\begin{equation}\label{eq_F}
 F(v)=\hbar(k_w+k_{st})\gamma \wp_e(v).
\end{equation}
This is exactly the result one would obtain by analogy with a two level system where the momentum transferred on each absorption/emission cycle is $\hbar (k_w+k_{st})$ (when averaged on several cycles) and where the number of cycles per unit time is the excited state scattering rate $\gamma$ times the probability to be in this excited state $\wp_e$. The introduction of the effective momentum $\mathbf{k}_{eff}=(\mathbf{k}_w+\mathbf{k}_{st})$ as the momentum transferred to the atoms is very significant. It demonstrates that the role of the coupling on the strong transition is not only to shorten the lifetime  of the metastable state $|m\rangle$, which would be independent on the relative direction of the two lasers. This sensitivity of the cooling process to the relative direction of propagation has been experimentally demonstrated in \cite{hendricks07}. 

$F(v)$ can be split in a constant force $F^0$, which acts on an immobile atom and in a velocity dependent force $F(v)-F^0$, responsible for the cooling process. For trapped particles, $F^0$ shifts the center of the trajectory by $F^0/m \omega_i^2$ and is not involved in the cooling. The cooling force, proportional to $\wp_e(v)- \wp_e(v=0)$  can be considered as a damping force $-\beta v$ only at  first order in $v$ ({\it i.e.} for $v \ll \Gamma_{eff}/k_x$)\cite{itano82b}.

 In the  approximation where $\Omega_w$ is the smallest relevant coupling parameter,  the analytic expression of the slope of $\wp_e$ with $v$ is  simple enough to  give an insight to the dependence of $\beta$ with the coupling parameters:
\begin{equation}\label{eq_beta}
\beta \simeq -\hbar (k_w+k_{st})^2 \gamma \wp_e^2(0)\frac{8\gamma^2\Delta_wk_w+8\left(4\Delta_w(\Delta_{st}+\Delta_w)-\Omega_{st}^2\right)\left(\Delta_w(k_{st}+k_w)+(\Delta_{st}+\Delta_w)k_w\right)   }{\Omega_{st}^2\Omega_w^2(k_w+k_{st})}.
\end{equation} 
First, one can check that for negative detuning $\Delta_{st}$  and negative relative detuning $\Delta_w-\delta_{LS}$, the $\beta$ coefficient is positive and damping effectively takes place.  Secondly, the particularity of the  regime  of large detunings $\Delta_{st}$ and low saturation ($\Omega_{st}^2\ll \Delta_{st}^2$)  for the strong transition is revealed by Eq.~\ref{eq_beta}. Indeed, in this limit, the detuning $\Delta_w$ (which is very close to $\delta_{LS}$), is negligible compared to $\Delta_{st}$ and then, only $k_w$ controls the slope of $\wp_e$ with $v$. It means that the velocity dependance of the force is proportional to $k_w v$, and that, at first order in $v$, the efficiency of the strong coupling is not reduced by the Doppler effect. Nevertheless, the relative direction of the two lasers remains relevant for the effective momentum $\mathbf{k}_{eff}$ transferred to the atoms  as the ratio between the sum and the difference of the two wavevectors $k_w$ and $k_{st}$ is equal to 11.9 for Ca$^+$, 4.4 for Sr$^+$ and 1.9 for Ba$^+$.

\begin{figure}[htb]
\begin{center}
\epsfig{file=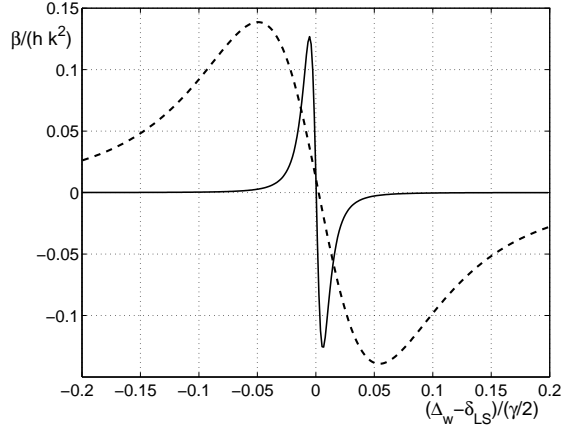, height=0.35\hsize} \caption{ Damping factor $\beta$  in units of $\hbar k^2$, versus the relative detuning of the weak transition from the light-shift.   Ca$^+$ wavelengths   are used ($\lambda_w=732$ nm and $\lambda_{st}=866$ nm) and $\gamma/2\pi=20$ MHz. The three curves are for different values of $\Omega_w/2\pi$: 0.1 MHz (solid line), 0.5 MHz (dashed line) and 1 MHz (dotted line), $\Delta_{st}/2\pi$ is fixed to $-10\gamma$ and $\Omega_{st}$ is chosen to obtain the maximum signal, according to Eq.\ref{eq_opt}.    } \label{fig_beta}
\end{center}
\end{figure}

When the light-shift becomes non-negligible compared to $\Delta_{st}$, the slope  becomes  steeper for copropagating lasers. However, the major effect of the relative direction of propagation remains  the size of the  effective momentum $\mathbf{k}_{eff}$.  In the following, we consider only the case where the two lasers are copropagating and call $k$ the maximum effective momentum.   Fig.~\ref{fig_beta} shows the damping coefficient $\beta$ for two different values of $\Omega_w$, in a regime where the effective linewidth depends on $\Omega_w$. The efficiency of the damping shows a maximum, which corresponds to a minimum of the cooling time, for detuning $\Delta_{w}-\delta_{LS}$ increasing with the linewidth of the signal. Numerical results show that, for fixed Rabi frequency on the weak transition $\Omega_w$, $\beta$ is maximised and so the cooling time minimised for $\Omega_{st}/\Delta_{st}$ close to its optimum defined to reach maximum fluorescence (see Eq.~\ref{eq_opt}). In practice, maximising the fluorescence rate also results in reducing the cooling time and this optimum condition was used for plots on Fig.~\ref{fig_beta}. The curves  show that, with this condition, the maxima of the damping are very close and that the advantage of an increase in $\Omega_w$ is mainly to broaden the frequency range where damping is efficient.

The temperature reached by Doppler cooling results from the equilibrium between the damping due to the momentum transfer and the diffusion induced by spontaneous emission. By conservation of energy, the scattered photons have the momentum $k$ and by analogy with two-level systems, the diffusion coefficient is \cite{gordon80}
\begin{equation}\label{eq_D}
D=\frac{1}{2}\hbar^2 k^2 \gamma \wp_e(0)
\end{equation}
and the temperature reached is $k_B T=D/\beta$.

\begin{figure}[htb]
\begin{center}
\epsfig{file=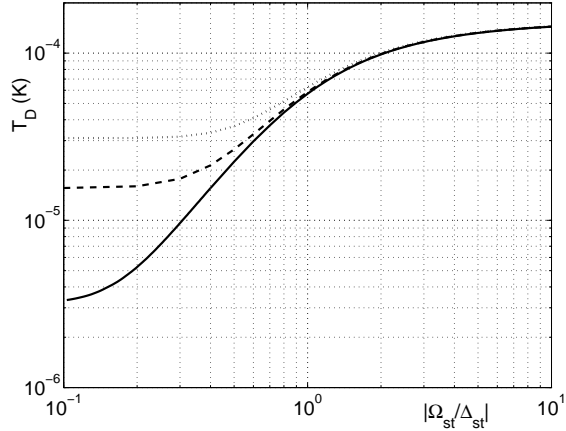, height=0.35\hsize} \caption{Doppler limit temperature versus $| \Omega_{st}/\Delta_{st}|$ for  Ca$^+$ wavelength ($\lambda_w=732$ nm and $\lambda_{st}=866$ nm) and $\gamma/2\pi=20$ MHz, $\beta_{eg}=1$.  For a given ratio $| \Omega_{st}/\Delta_{st}|$ the curves are identical  for different $\Delta_{st}$. The different curves are for different values of $\Omega_w/2\pi$: 0.1 MHz (solid line), 0.5 MHz (dashed line) and 1 MHz (dotted line).    } \label{fig_TD}
\end{center}
\end{figure}

\begin{figure}[htb]
\begin{center}
\epsfig{file=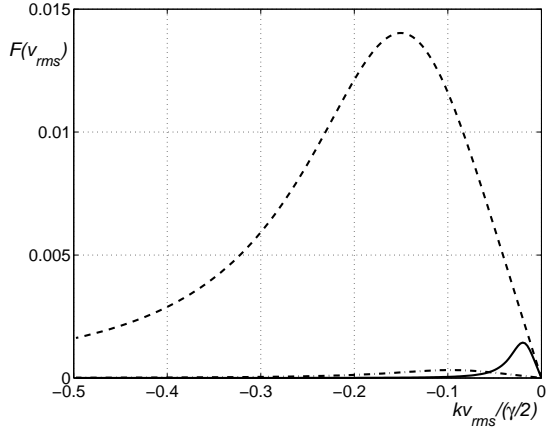, height=0.35\hsize} \caption{ Effective force $\overline{F}(v_{rms})$ in units of $\hbar k \gamma$ versus $kv_{rms}/(\gamma/2)$ for  Ca$^+$ wavelength ($\lambda_w=732$ nm and $\lambda_{st}=866$ nm) and $\gamma/2\pi=20$ MHz, $\beta_{eg}=1$.   The different curves are for different values of $\Omega_w/2\pi$: 0.1 MHz (solid line and dot-dashed line) and 1 MHz (dashed line). $\Delta_{st}$ is fixed to $-10\gamma$ and $\Omega_{st}$ is chosen to obtain the maximum signal, according to Eq.\ref{eq_opt}, only for the curves plotted with the solid line and the dashed line: $ \Omega_{st}=34$ MHz (solid line), 106 MHz (dashed and dot-dashed line).  $\Delta_w=\delta_{LS}-\Gamma_{eff}/2$.   } \label{fig_capture}
\end{center}
\end{figure}

Numerical calculations show that the minimum temperature $T_D$ is reached for $\Delta_w-\delta_{LS}=-\Gamma_{eff}/2$ like expected by analogy with a two level system. They also show that the relevant parameter for the strong coupling is the ratio $| \Omega_{st}/\Delta_{st} |$ since for identical ratio, different values of $\Delta_{st}$ or $\Omega_{st}$ give the same results. The minimum temperature $T_D$ is plotted on Fig.~\ref{fig_TD} versus $|\Omega_{st}/\Delta_{st}|$, for the wavelengths of Ca$^{+}$ and with copropagating lasers. 
 Like the effective linewidth, the limit temperature can be adjusted continuously down to few $\mu$K  by the choice of  $\Omega_{st}/\Delta_{st}$, the lowest temperature being controlled by $\Omega_w$.  As a comparison, the limit temperature reached by Doppler cooling on the allowed  dipole $|g\rangle \to |e\rangle$ transition on   Ca$^+$ ions is 0.55 mK. Contrary to a two-level system, the   Doppler limit temperature can be lowered, but to the  detriment of the fluorescence rate. Nevertheless, as low temperatures can be reached by continously reducing the effective linewidth, the cooling process can be started with a high damping factor, resulting in a total cooling time shorter than if the cooling process would be made permanently on a very narrow transition.  As for the capture range, defined by the velocity for which the cooling force is maximal, numerical results confirm the intuitive idea that it increases with the effective linewidth $\Gamma_{eff}$, which in practice, increase with $| \Omega_{st}/\Delta_{st}|$ and/or $\Omega_w$ (see part \ref{sec_width}). It is possible to reach a broad capture range at the same time as keeping a high damping factor and fluorescence rate by incresing $| \Omega_{st}/\Delta_{st}|$ and obeying the optimum condition defined in \ref{sec_popP} (Eq.~\ref{eq_opt}). This is shown on Fig.~\ref{fig_capture} where the dependence of the effective force $\overline{F}_{rms}$  with the root mean square velocity is plotted for the condition leading to the minimum Doppler limit, $\Delta_w=\delta_{LS}-\Gamma_{eff}/2$. Within the assumption defined at the beginning of this section, this effective force can be approximated by  $F(v_{rms})-F(-v_{rms})$.

 The lowest temperatures shown on Fig.~\ref{fig_TD} have only a theoretical meaning as, in practice, for reaching these temperatures, the effective linewidth of the system must be reduced and reach  values smaller than the recoil and the oscillation frequencies. Then, to make predictions, other mechanisms than the Doppler cooling  have to be taken into account, as the well known sideband cooling  which will not be developed here but for which efficiency has been experimentally demonstrated \cite{diedrich89, roos99}.

\section{Conclusions}\label{Sec:Conclusions}
We have studied a two-step photoexcitation process in a ladder-system with a metastable intermediate state, as can  be found {\it e.g.} in alkaline-earth atoms and in alkali-like ions. In case of Ca$^+$, Sr$^+$ and Ba$^+$, the two excitation wavelengths belong to the red or infrared domain whereas the scattered photons lie in the blue domain, which offers the possibility of fluorescence detection without stray light background. Our numerical and analytical studies show that this excitation process can be a powerful alternative for the detection of ions or readout of their internal state. Indeed,  the number of photons can easily reached a tenth of what is expected from direct excitation on a dipole allowed transition and the absence of background signal due to scattered light can compensate this lower signal. Furthermore, we demonstrate that two-step Doppler cooling can efficiently take place and  that resolution of oscillation sidebands is not required to reach temperatures lower than 100 $\mu$K. The width of the resonance profile and the Doppler limit temperature can be varied continuously by tuning the power and/or detuning on the dipole allowed transition. This cooling mechanism can be of great interest for traps with low  oscillation frequencies or with geometry leading to a high stray light level like microtraps.

\begin{acknowledgments} C.C. is grateful to Jean Dalibard for precious advise and M.D. acknowledges the hospitality of laboratory PIIM and thanks Universit\'e de Provence for his grant as a guest professor. The present collaboration has benefitted from financial support by the French Embassy in Denmark.
   \end{acknowledgments}

%\bibliography{/home/caroline/biblioCC/bibliotrap}

\end{document}